\documentclass[a4paper,11pt]{article}

\usepackage{jheppub} 
\usepackage[utf8]{inputenc}

\usepackage[force]{feynmp-auto}

\renewcommand{\Re}{\operatorname{Re}}

\newsavebox{\tmp}

\title{Meromorphic solutions of recurrence relations and DRA method for multicomponent master integrals}

\author{Roman N. Lee}
\author{and Kirill T. Mingulov}

\affiliation{The Budker Institute of Nuclear Physics,\\630090, Novosibirsk}
\emailAdd{r.n.lee@inp.nsk.su}
\emailAdd{k.t.mingulov@inp.nsk.su}

\abstract{
	We formulate a method to find the meromorphic solutions of higher-order recurrence relations in the form of the sum over poles with coefficients defined recursively. Several explicit examples of the application of this technique are given. The main advantage of the described approach is that the analytical properties of the solutions are very clear (the position of poles is explicit, the behavior at infinity can be easily determined). These are exactly the properties that are required for the application of the multiloop calculation method based on dimensional recurrence relations and analyticity (the DRA method).}

\begin{document} 
\maketitle
\flushbottom

\section{Introduction}
\label{sec:intro}

The ability to perform multiloop calculations is very essential for 
obtaining high-precision theoretical predictions in particle physics. Such predictions are, in particular, indispensable for the ongoing searches for New Physics. Since 1980s the field of multiloop calculations has experienced an explosive development in terms of the available methods and tools. Two major technical achievements in this region are the integration-by-parts reduction \cite{ChetTka1981,Tkachov1981} and differential equations method \cite{Remiddi1997,Kotikov1991b}. Thanks to these two techniques, the present frontiers of the multiloop calculations reside somewhere close to $\mathrm{NNLO}$ calculation of the differential cross sections with up to $4$ or $5$ particles involved (i.e., two loops and $4$ parameters). However, some quantities, which depend only on one invariant, deserve and allow for the $\mathrm{N}^3\mathrm{LO}$ and $\mathrm{N}^4\mathrm{LO}$ accuracy. The differential equations can not help, at least directly, in this case. Probably, the most celebrated example are perturbative contributions to the anomalous magnetic moment of the electron and muon. Due to the efforts of Kinoshita's group (see Ref.~\cite{Aoyama2015} and references therein), the QED contributions to electron $g-2$ were known numerically up to the four-loop accuracy, and only recently these results have been independently verified  by Laporta \cite{Laporta2017}. In fact the approach of Ref.~\cite{Laporta2017} would lead to completely analytical form of the four-loop QED contribution to electron $g-2$ if we understood better which transcendental numbers might enter the final expression. This is, probably, another great achievement of contemporary multiloop methods --- learning how to guess the analytical results from the high-precision numerical ones. 

One of the methods which may be used for the calculation of the one-scale integrals is the method of Ref.~\cite{Laporta2000} based on the difference equations with respect to the power of one of the massive propagators. Another natural idea is to use the recurrence relations with respect to the space-time dimension $d$, \cite{Tarasov1996}. In Ref.~\cite{Lee2010} the DRA method was formulated which uses the dimensional recurrence relations and the analyticity of the integrals as functions of $d$. The key idea of the DRA approach is to use the analytical properties of the integrals in order to fix the form of the homogeneous solutions up to several constants which should be determined by other methods (e.g. from Mellin-Barnes representation). The results of the DRA method have the form of multiple convergent sums with factorized summand. The latter property allows for their fast high-precision evaluation. The corresponding algorithms were implemented in the \texttt{SummerTime} program \cite{LeeMingulov:2016:SummerTime}, which allows one to obtain the $\epsilon$ expansion of the integrals near any dimensionality $d$ with high-precision numerical coefficients. This method was successfully applied in many physical calculations \cite{Lee2010a,LeeSmSm2010a,LeeSmi2010,LeeSmSm2011,LeeTer2010,Lee2011e,LeeSmirnov2012,LeeMarquardSmirnovSmirnovSteinhauser2013,LeeSmirnov2016,LeeSmirnovSmirnovSteinhauser2016}.

In order to apply the DRA method it is necessary to construct the homogeneous solutions with known analytical properties. While for the first-order recurrence relations this is a trivial task, for the higher-order recurrence relations this is not so. In Ref.~\cite{LeeSmirnov2012} the homogeneous solutions of the second-order differential equations were obtained from the explicit evaluation of maximally cut integrals. This approach related a concrete example of the integrals and can be hardly generalized. Therefore, the application of the DRA method to the topologies containing the sectors with several master integrals (multicomponent master integrals in terminology of Ref.~\cite{LeeSmirnov2012}) is complicated by the necessity to construct the homogeneous solution of the higher-order recurrence relations with known analytical properties.

To explain the main goal of this paper let us consider the following example. Suppose that we want to find the meromorphic function $f(\nu)$ ($\nu\in \mathbb{C}$) which obeys the first-order recurrence relation
\begin{equation}
P(\nu)f(\nu)+Q(\nu)f(\nu+1)=0\,,
\end{equation}
where $P$ and $Q$ are some polynomials. This is a simple problem: we can take, e.g., the function
\begin{equation}
f(\nu)=\left(-\frac{p}{q}\right)^\nu\frac{\prod_{k} \Gamma(\nu-a_k)}{\prod_{l} \Gamma(\nu-b_l)}\,,
\end{equation}
where $p$, $q$, $a_k$, and $b_l$ are determined by the factorization of the polynomials $P$ and $Q$:
\begin{equation}
P(\nu)=p \prod_{k}(\nu-a_k)\,,\quad Q(\nu)= q\prod_{l} (\nu-b_l)\,.
\end{equation}

Suppose now that we have the second-order recurrence relation, e.g.
\begin{equation}
\label{eq:2nd-order-rec}
P(\nu)f(\nu)+Q(\nu)f(\nu+1)+R(\nu)f(\nu+2)=0\,.
\end{equation}
Then the task of finding the meromorphic solution is not simple anymore.
In principle, one can try some integral transformations, e.g.\ the Mellin transformation, in order to turn Eq.~\eqref{eq:2nd-order-rec} into the differential equation. However, the order of this differential equation and the number of its singular points grows rapidly with the degree of the polynomial coefficients $P$, $Q$, and $R$ which makes this approach impractical in multiloop calculations.
The main goal of this paper is to suggest the alternative approach based on searching the solutions in the form of the sum over poles with coefficient defined recursively. We will show on several examples that the formulated approach can be successfully used for finding the general homogeneous solutions.
In the parallel paper \cite{DREAM}, we present the package \texttt{DREAM} which allows one to automatically construct the inhomogeneous solutions.
In that paper, we use this package and the homogeneous solutions found in the present paper to calculate the corresponding multiloop integrals.

\section{Meromorphic solutions of recurrence relations}

Let us consider the $n$-th order homogeneous recurrence relation
\begin{equation}\label{eq:recurrence}
\mathcal{P}f(\nu)=\sum_{k=0}^n p_k(\nu+k)f(\nu+k)=0\,,
\end{equation}
where $p_k(\nu)$ are some given polynomials and we shifted the arguments of $p_k$ for convenience of further considerations. We will assume that $\deg p_0=\deg p_n=N$,  $\deg p_i\leqslant N$. We are interested in $n$ independent solutions $f_1(\nu),\ldots,f_n(\nu)$, each being a meromorphic function of the complex variable $\nu$.
Then the general solution can be represented as 
\begin{equation}	
f(\nu)=\sum_{k=1}^{n}\omega_k(\nu) f_k(\nu)\,,
\end{equation}
where $\omega_1(\nu),\ldots,\omega_n(\nu)$ are arbitrary periodic functions with period $1$.

Note that for $n=1$ we can explicitly write such a solution as	
\begin{equation}
f_1(\nu)=\left(-\frac{p_{1N}}{p_{0N}}\right)^\nu\frac{\prod_{i=1}^{N} \Gamma(\nu+1-\nu_{1i}) }{\prod_{i=1}^{N} \Gamma(\nu-\nu_{0i}) }\,,
\end{equation}
where $p_{kN}$ is the leading coefficient of  $p_{k}(\nu)= p_{kN}\nu^N+\ldots$, and $\nu_{ki}$ are the zeros of $p_{k}(\nu)$. When $n>1$ the problem of finding the meromorphic solutions is much more difficult. 

Let us search for the solution in the form
\begin{equation}\label{eq:solutionForm}
f(\nu)=\lambda^\nu\sum_{\sigma\in{S+\mathbb{Z}}} \frac{c(\sigma)\lambda^{-\sigma}}{\nu-\sigma}\,,
\end{equation}
where the coefficients $c(\sigma)$ satisfy the recurrence
\begin{equation}\label{eq:c_recurrence}
	\sum_{k=0}^np_k(\sigma+k)c(\sigma+k)=0\qquad (\sigma\in  {S+\mathbb{Z}})\,,
\end{equation}
 $\lambda$ is some number, and $S=\{\sigma_1,\ldots, \sigma_m\}$ is some finite set with no resonances, i.e. the difference between any two distinct elements of $S$ is non-integer, $\sigma_i-\sigma_j\not\in \mathbb{Z}$. 
For the moment we assume that the sum over $\sigma$ converges absolutely for any $\nu\in \mathbb{C}\backslash (S+\mathbb{Z})$.
 
Let us substitute the form \eqref{eq:solutionForm} in the left-hand side of \eqref{eq:recurrence}. We have
\begin{equation}\label{eq:recurrencesubst}
\lambda^{-\nu} \mathcal{P}f(\nu) = \sum_{k=0}^n p_k(\nu+k)\lambda^k\sum_{\sigma\in  {S+\mathbb{Z}}} \frac{c(\sigma)\lambda^{-\sigma}}{\nu+k-\sigma}\,.
\end{equation}
It is easy to check that Eq.~\eqref{eq:recurrencesubst} defines an entire function of $\nu$, thanks to Eq.~\eqref{eq:c_recurrence}. Indeed, the poles are only possible at $\nu\in S+\mathbb{Z}$, but taking the residue and using the recurrence relation \eqref{eq:c_recurrence} we see that those poles cancel.

Now we will determine the conditions at which the entire function \eqref{eq:recurrencesubst} vanishes. Let us write $\frac{p_k(\nu+k)}{\nu+k-\sigma}=\frac{p_k(\sigma)}{\nu+k-\sigma}+q_k(\nu+k,\sigma)$, where
\begin{equation}
	q_k(\nu,\sigma)=\frac{p_k(\nu)-p_k(\sigma)}{\nu-\sigma}\,.
\end{equation}
Then we have 
\begin{equation}\label{eq:recurrence1}
 \lambda^{-\nu}\mathcal{P}f(\nu) = \sum_{\sigma} c(\sigma)\lambda^{-\sigma}Q(\nu,\sigma) + \sum_{k=0}^n \sum_{\sigma\in  {S+\mathbb{Z}}} p_k(\sigma)\lambda^{k-\sigma}\frac{c(\sigma)}{\nu+k-\sigma}\,,
\end{equation}
where 
\begin{equation}\label{eq:Q}
	Q(\nu,\sigma)=\sum_{k=0}^n \lambda^{k}q_k(\nu+k,\sigma)\stackrel{\text{def}}{=}\sum_{l=0}^{N-1} Q_{l}(\sigma)\nu^l
\end{equation} is the polynomial of the degree $N-1$, the last equality in \eqref{eq:Q} defines the polynomials $Q_l(\sigma)$.
The second term in the right-hand side of Eq.~\eqref{eq:recurrence1} can be easily shown to vanish by shifting $\sigma\to \sigma+k$: 
\begin{multline}\label{eq:term2}
\sum_{k=0}^n \sum_{\sigma\in  {S+\mathbb{Z}}} p_k(\sigma)\lambda^{k-\sigma}\frac{c(\sigma)}{\nu+k-\sigma}
=\sum_{k=0}^n \sum_{\sigma\in  {S+\mathbb{Z}}} p_k(\sigma+k)\lambda^{-\sigma}\frac{c(\sigma+k)}{\nu-\sigma}\\
= \sum_{\sigma\in  {S+\mathbb{Z}}} \lambda^{-\sigma}\frac{\sum_{k=0}^np_k(\sigma+k)c(\sigma+k)}{\nu-\sigma}=0\,.
\end{multline}
Therefore, we have to require that the following $N$ conditions hold:
\begin{align}
\sum_{\sigma\in  {S+\mathbb{Z}}} c(\sigma)\lambda^{-\sigma}Q_0(\sigma)&=0\label{eq:constr1}\\
\vdots\nonumber\\
\sum_{\sigma\in  {S+\mathbb{Z}}} c(\sigma)\lambda^{-\sigma}Q_{N-2}(\sigma)&=0\label{eq:constr2}\\
\sum_{\sigma\in  {S+\mathbb{Z}}} c(\sigma)\lambda^{-\sigma}Q_{N-1}(\sigma)&=0.
\end{align}

The last condition is somewhat special, since $Q_{N-1}$ does not depend on $\sigma$. Let $p_k(\nu)=p_{kN}\nu^N+\ldots$ Then it is easy to see that $Q_{N-1}=\sum_{k=0}^n p_{kN}\lambda^k$.
If $\lambda$ is a root of the characteristic equation
\begin{equation}\label{eq:char}
\sum_{k=0}^n p_{kN} \lambda^k=0,
\end{equation}
we have $Q_{N-1}=0$. 

From now on we will assume assume that $\lambda$ is one of the roots of Eq.~\eqref{eq:char}. Therefore, we have $N-1$ equations \eqref{eq:constr1}--\eqref{eq:constr2}. Suppose that we fix somehow the set $S$. Let us count the number of free parameters which we might tune to secure  constraints \eqref{eq:constr1}--\eqref{eq:constr2}. Since $c(\sigma)$ satisfy the $n$-th degree recurrence relation \eqref{eq:c_recurrence}, we might fix arbitrarily the coefficients $c(\sigma_l),\,c(\sigma_l+1),\ldots,c(\sigma_l+n-1)$ for each $\sigma_l\in S$. Therefore, we have $n\times m$ parameters ($m=|S|$). A naive counting would be that when $n\times m\geqslant N$, we have a nontrivial solution of \eqref{eq:constr1}--\eqref{eq:constr2}. However, in general the convergence requirements should also be treated properly. Suppose that we fix $c(\sigma_l),\,c(\sigma_l+1),\ldots,c(\sigma_l+n-1)$ arbitrarily. Then the asymptotics of the sequence $c(\sigma_l+k)$ is likely to behave as $\lambda_{\mathrm{max}}^k$ when $k\to +\infty$ and $\lambda_{\mathrm{min}}^k$ when $k\to -\infty$. Here $\lambda_{\mathrm{max}}$ ($\lambda_{\mathrm{min}}$) denote the roots of the characteristic equation \eqref{eq:char} with maximal (minimal) absolute value, respectively. Therefore, whatever $\lambda$ we take, the sums in \eqref{eq:constr1}--\eqref{eq:constr2} are likely to diverge at the upper and/or at the lower limit. Fortunately, there is a way out of these convergence problems. Let us fix $S=\{\sigma_1,\ldots \sigma_N\}$, where $\sigma_1,\ldots \sigma_N$ are the zeros of the polynomial $p_n(\sigma)$. Then putting $c(\sigma_l+n)=0$ for all negative $n$ and $c(\sigma_l)=\mathrm{const}\neq0$ is obviously consistent with the recurrence relations \eqref{eq:c_recurrence}. In other words, for this specific choice of $S$ the summation limits in \eqref{eq:constr1}--\eqref{eq:constr2} are restricted from below, $\sigma\in S+\mathbb{Z}_+$, where $\mathbb{Z}_+=\{0,1,2,\ldots\}$. Therefore, we should only care for the convergence at the upper limit. The choice $\lambda=\lambda_{\mathrm{max}}$ eliminates at least exponential divergence and we shall see on the specific examples that the sums converge. Then, $N-1$ equations \eqref{eq:constr1}--\eqref{eq:constr2} form a linear system for $N$ variables $c(\sigma_1),\ldots,c(\sigma_N)$ from which these variables can be determined up to the arbitrary common factor.

Therefore, we have a receipt to find at least one meromorphic solution of the recurrence relation \eqref{eq:recurrence}. In fact, setting $S$ to be the set of zeros of $p_0(\sigma)$ and repeating the similar analysis, we may find the second solution for which the summation goes downwards. For the second-order recurrences these two solutions constitute a complete set, apart from possible degeneracy.

In the next Section we demonstrate that the presented approach can be successfully applied to very nontrivial multiloop integrals.

\section{Examples}

\subsection*{Three-loop massive sunrise integral on the pseudo-threshold}
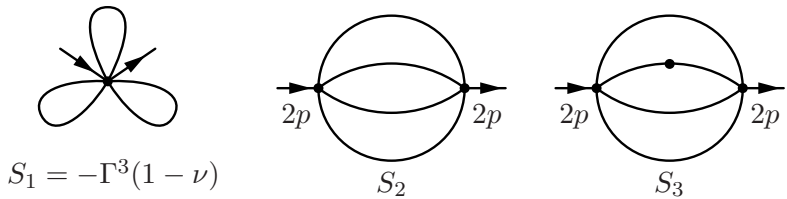
\begin{figure}[h]
	\unitlength = 0.5mm
	\vspace{5mm} 	    
	\centering
	\begin{fmffile}{mis0}
		\parbox{3cm}{
			\centering
			\raisebox{-5mm}{\begin{fmfgraph}(25,25)
				\fmfset{arrow_len}{3mm}
				\fmfset{dot_size}{1mm}
				\fmftop{v1,v2}
				\fmfbottom{v3}
				\fmfdot{v}
				\fmf{fermion,tension=0.8}{v1,v,v2} 
				\fmf{phantom,tension=1}{v3,v} 
				\fmf{plain,left=45,tension=0.5}{v,v} 
				\fmf{plain,left=135,tension=0.5}{v,v} 
				\fmf{plain,left=-135,tension=0.5}{v,v}
				\fmffreeze
			\end{fmfgraph}}\vspace{3mm}\\ 
			\centering $S_1=-\Gamma^3(1-\nu)$} \hspace{4mm}
		\parbox{3cm}{
			\centering
			\begin{fmfgraph*}(60,30)
				\fmfset{arrow_len}{3mm}
				\fmfset{dot_size}{1mm}
				\fmfleft{l1}
				\fmfright{r1}
				\fmf{fermion,label=$2p$}{l1,v1} 
				\fmf{fermion,label=$2p$}{v2,r1} 
				\fmf{plain,tension=0.07,left=1}{v1,v2,v1} 
				\fmf{plain,tension=0.07,left=0.34}{v1,v2,v1} 
				\fmfdot{v1,v2}
			\end{fmfgraph*}\vspace{1mm}\\ 
			\centering $S_2$} \hspace{4mm} 
		\parbox{3cm}{
			\centering
			\begin{fmfgraph*}(60,30)
				\fmfset{arrow_len}{3mm}
				\fmfset{dot_size}{1mm}
				\fmfleft{l1}
				\fmfright{r1}
				\fmftop{t}
				\fmfbottom{b}
				\fmf{phantom,tension=2.5}{t,d}
				\fmf{phantom}{d,b}
				\fmfdot{d}				
				\fmf{fermion,label=$2p$}{l1,v1} 
				\fmf{fermion,label=$2p$}{v2,r1} 
				\fmf{plain,tension=0.07,left=1}{v1,v2,v1} 
				\fmf{plain,tension=0.07,left=0.34}{v1,v2,v1} 
				\fmfdot{v1,v2}
			\end{fmfgraph*}\vspace{1mm}\\ 
			\centering $S_3$} 
	\end{fmffile}
	\caption{Master integrals. Solid lines correspond to the propagators $1/(k^2-1)$.}\label{fig:mis0}	
	\vspace{5mm} 	    
\end{figure}

Let us consider the three-loop  massive sunrise topology on the threshold. There are three master integrals depicted in Fig.~\ref{fig:mis0}.
The first master integral is trivial, $S_1=-\Gamma^3(1-\nu)$, and the last two satisfy a coupled system of dimensional recurrence relations:
\begin{align}
S_2(\nu+1)=&-\frac{8 \left(66 \nu ^2-106 \nu
	+43\right) S_1(\nu )}{(3 \nu -2)_2(4 \nu -3)_3}
	+\frac{4\left(324 \nu ^2-738 \nu +425\right) S_2(\nu
	)}{(3 \nu -2)_2(4 \nu
	-3)_3}\nonumber\\
	&-\frac{240 (6 \nu -7) S_3(\nu
	)}{(3 \nu -2)_2(4 \nu -3)_3}, \\
S_3(\nu+1)=&-\frac{2\left(42 \nu ^2-67
	\nu +27\right) S_1(\nu )}{(3 \nu
	-2)_2(4 \nu -3)_2}+\frac{6 \left(30
	\nu ^2-70 \nu +41\right) S_2(\nu )}{(3 \nu
	-2)_2(4 \nu -3)_2}\nonumber\\
	&-\frac{36 (6 \nu -7) S_3(\nu )}{(3 \nu
		-2)_2(4 \nu -3)_2}\,.
\end{align} 
Here and below $\nu=d/2$, where $d$ is space-time dimension, and $\alpha_n=\alpha(\alpha+1)\cdots(\alpha+n-1)$ is Pochhammer symbol. 
These equations lead to the second-order recurrence relation for $S_2$:
\begin{multline}\label{eq:S2recurrence}
-\frac{3 (6 \nu -5) S_2(\nu )}{ (2 \nu
	-1) (3 \nu -2)_2}
-\frac{\left(54
	\nu ^2+1\right) S_2(\nu +1)}{3 (6 \nu
	-1)}
+\frac{
	(3 \nu +1)_2(4 \nu +1)_3
	S_2(\nu +2)}{24 (6 \nu -1)}\\
=\frac{\left(432 \nu ^4-180 \nu ^3-46 \nu ^2+11
	\nu +3\right) \Gamma (-\nu )^3}{3 (2 \nu
	-1) (3 \nu -1) (6 \nu -1)}
\end{multline}
The homogeneous part of Eq.~\eqref{eq:T4recurrence} is not of the form of Eq.~\eqref{eq:recurrence} because the degrees of the polynomial coefficients do not satisfy the conditions $\deg p_0=\deg p_n=N$,  $\deg p_i\leqslant N$. However, we can easily fix this by passing to a new function $s_2$, e.g. with
\begin{equation}
S_2(\nu)=F(\nu)s_2(\nu)\,,
\end{equation}
where 
\begin{equation}
F(\nu)=\Gamma \left(\tfrac{4}{3}-\nu \right) \Gamma
\left(\tfrac{3}{2}-\nu \right) \Gamma
\left(\tfrac{5}{3}-\nu \right)
\,.
\end{equation}
Substituting this in Eq.~\eqref{eq:S2recurrence}, we have
\begin{gather}\label{eq:s2recurrence}
\sum_{k=0}^2 p_k(\nu+k)s_2(\nu+k)	= r(\nu)\,,
\end{gather}
where
\begin{align}
p_0(\nu)=&-(6 \nu -5) (6 \nu -1)\,,\nonumber\\
p_1(\nu)=&2 \left(54 (\nu -1)^2+1\right)\,,\nonumber\\
p_2(\nu)=&9 (4 \nu -7) (4 \nu -5)\,,\nonumber\\
r(\nu) =&-\frac{2 \left(432 \nu ^4-180 \nu ^3-46 \nu
	^2+11 \nu +3\right) \Gamma (-\nu )^3}{(2 \nu
	-1) (3 \nu -1)F(\nu+1)}
\,.\label{eq:p&r}
\end{align}	
We now apply the method described in the previous section to find the homogeneous solution of Eq.~\eqref{eq:s2recurrence}, i.e., solution of
\begin{equation}\label{eq:homo}
\sum_{k=0}^2 p_k(\nu+k)s_{2h}(\nu+k) = 0\,.
\end{equation}
The characteristic equation $144 \lambda ^2+108 \lambda -36=0$ has the solutions
\begin{equation}
\lambda_1=-1\,,\quad \lambda_2=\frac14\,.
\end{equation}
The zeros of $p_2(\nu)$ are
\begin{equation}
\nu_1=\frac54\,,\quad \nu_2=\frac74\,.
\end{equation}
Here we note that $\lambda_1=-1$ which determines the behavior of the homogeneous solution at $\nu\to +\infty$ is negative. Therefore it is convenient to pass to the function $\tilde{s}_2(\nu)=a(\nu)s_2(\nu)$, where $a(\nu)=-a(\nu+1)$ is some anti-periodic function which we choose as $a(\nu)=\sin(\pi\nu)$. Such a substitution flips the signs of $p_1(\nu)$ and $\lambda_{1,2}$. Then, according to the prescriptions of the previous section, we search for the solution in the form
\begin{equation}\label{eq:s2sol}
s_{2h}^{(1)}(\nu)=\frac{1}{\sin(\pi\nu)}\sum _{n=0}^{\infty } \left[\frac{a_n}{\nu -\nu_1-n}+ \frac{b_n}{\nu -\nu_2 - n}\right]\,,
\end{equation}
where we have taken into account that $-\lambda_1=1$ and the sequences $\{a_0,a_1,a_2,\ldots\}$ and $\{a_0,b_1,b_2,\ldots\}$ are defined recursively:
\begin{gather}
a_{-1}=0,\ a_{0}=1,\ a_{n\geq 1}=\frac{432 n^2-648 n+251}{288 n (2 n-1)}a_{n-1}+\frac{(12 n-19) (12 n-11) }{288 n (2 n-1)}a_{n-2}\,,\nonumber\\
b_{-1}=0,\ 
 b_{n\geq 1}=\frac{432 n^2-216 n+35}{288 n (2 n+1)}b_{n-1}+\frac{(12 n-13) (12 n-5)}{288 n (2 n+1)}b_{n-2}\,.\label{eq:ab-seq}
\end{gather} 

In order to determine the unfixed coefficient $b_0$ in Eq.~\eqref{eq:t4sol}, we calculate the polynomial $Q(\nu,\sigma)$, Eq.~\eqref{eq:Q}. It appears that $Q=0$ and one might think that Eq.~\eqref{eq:s2sol} is a solution for arbitrary $b_0$. There is, however, a convergence issue which we have to take care of. Namely, we tacitly assumed that the sum $\sum_{\sigma\in  {S+\mathbb{Z}}} p_k(\sigma)\lambda^{k-\sigma}\frac{c(\sigma)}{\nu+k-\sigma}$ in the first expression of Eq.~\eqref{eq:term2} converges. Specialization to the case under consideration is that
\begin{equation}\label{eq:sum}
\sum_{n=0}^{\infty}\left[ p_k(\nu_1+n)\frac{a_n}{\nu+k-\nu_1-n}+p_k(\nu_2+n)\frac{b_n}{\nu+k-\nu_2-n}\right]
\end{equation}
should converge. The asymptotics of the sequences $a_n$ and $b_n$ can be found along the lines of Ref.~\cite{Tulyakov:2011}. We have
\begin{equation}
a_n,b_n\sim \frac{1}{n}\,,
\end{equation}
therefore, a necessary condition of the convergence of \eqref{eq:sum} reads 
\begin{equation}
\lim_{n\to\infty} n (a_n+b_n) = 0\,.
\end{equation}
from which we have
\begin{equation}
b_0=-0.32977278946\ldots
\end{equation}
Using some heuristic conjectures and \texttt{PSLQ} algorithm \cite{FergBai1991}, we find
\begin{equation}\label{eq:b0}
b_0\stackrel{1000}{=}-\frac{5 \Gamma \left(\frac{3}{4}\right)^2}{\sqrt{3} \Gamma \left(\frac{1}{4}\right)^2}\,.
\end{equation}
Note that the sum in Eq.~\eqref{eq:sum} diverges logarithmically even after we put $b_0$ to the value \eqref{eq:b0}, however this is sufficient for the consistency of summation variable shifts and changing the summation order used to prove Eq.~ \eqref{eq:term2}. Therefore, explicitly, we have the following homogeneous solution
\begin{equation}\label{eq:s2h1}
s_{2h}^{(1)}(\nu)=\frac{1}{\sin(\pi\nu)}\sum _{n=0}^{\infty } \left[\frac{a_n}{\nu -5/4-n}+ \frac{ b_n}{\nu -7/4 - n}\right]\,,
\end{equation}
where $a_n$ and $b_n$ are defined by Eqs.~\eqref{eq:ab-seq} and \eqref{eq:b0}.

The second homogeneous solution can be found in a similar way by taking $S$ to be the set of zeros of the trailing coefficient $p_0(x)$, i.e., $S=\left\{\frac16,\frac56\right\}$. However it is simpler to use the hidden symmetry of Eq.~\eqref{eq:homo}. Namely, Eq.~\eqref{eq:homo} is invariant under the replacement
$s_{2h}(\nu)\to \frac{\sin (\pi  \nu )\cos (2\pi  \nu )\Gamma \left(\frac{7}{4}-\nu \right) \Gamma \left(\frac{5}{4}-\nu \right)s_{2h}(2-\nu)}{4^{\nu }  \Gamma \left(\frac{7}{6}-\nu \right) \Gamma \left(\frac{11}{6}-\nu \right)}$ followed by $	\nu\to 2-\nu$. Therefore, we may write the second solution as
\begin{equation}\label{eq:s2h2}
s_{2h}^{(2)}(\nu)= \frac{ \sin (\pi  \nu )\cos (2\pi  \nu )\Gamma \left(\frac{7}{4}-\nu \right) \Gamma \left(\frac{5}{4}-\nu \right)s_{2h}^{(1)}(2-\nu)}{4^{\nu }\Gamma \left(\frac{7}{6}-\nu \right) \Gamma \left(\frac{11}{6}-\nu \right)}\,.
\end{equation}
It is easy to check numerically that $s_{2h}^{(2)}(\nu)$ is independent of $s_{2h}^{(1)}(\nu)$, i.e., that their ratio is not a periodic function. 
\subsubsection*{Fixing periodic factors}
Let us explain now how the homogeneous solutions found can be used within the DRA method and allow one to fix the form of the result for $s_2$ (and for $S_2$). We write the general solution of \eqref{eq:s2recurrence} in the form 
\begin{equation}\label{eq:gen}
{s_2(\nu+1)\choose s_2(\nu)}= 
\left[
\begin{array}{cc}
s_{2h}^{(1)}(\nu+1)&s_{2h}^{(2)}(\nu+1)\\
s_{2h}^{(1)}(\nu)&s_{2h}^{(2)}(\nu)
\end{array}
\right]
{\omega_1(\nu)\choose \omega_2(\nu)}+{s_{2ih}(\nu+1)\choose s_{2ih}(\nu)}\,,
\end{equation}
where $\omega_{1,2}(\nu)$ are arbitrary periodic functions and $t_{4ih}(\nu)$ is the special solution of the inhomogeneous equation. The latter can be written as
\begin{equation}\label{eq:s2ih}
{s_{2ih}(\nu+1)\choose s_{2ih}(\nu)}=R(\nu)+\sum_{n=0}^{\infty}A(\nu)\ldots A(\nu-n)R(\nu-n-1)\,,
\end{equation}
where 
\begin{equation}\label{eq:AR}
A(\nu)=
\left(
\begin{array}{cc}
-\frac{p_1(\nu)}{p_2(\nu+1)}&-\frac{p_0(\nu-1)}{p_2(\nu+1)}\\
1&0
\end{array}
\right)\,,\quad R(\nu)= {\frac{r(\nu-1)}{p_2(\nu+1)}\choose 0}\,,
\end{equation}
and $p_0,\, p_1,\, p_2,\, r$ are defined in Eq.~\eqref{eq:p&r}. Note that $s_{2ih}(\nu)$ does not have poles in the region $\Re\nu<1$. Now, according to the standard prescription of the DRA method, we rewrite Eq.~\eqref{eq:gen} as
\begin{equation}\label{eq:om}
{\omega_1(\nu)\choose \omega_2(\nu)}= T(\nu)
{s_{2}(\nu+1)-s_{2ih}(\nu+1)\choose s_{2}(\nu)- s_{2ih}(\nu)}
\,,
\end{equation}
where  
\begin{align}
T(\nu)=&\left[
\begin{array}{cc}
s_{2h}^{(1)}(\nu+1)&s_{2h}^{(2)}(\nu+1)\\
s_{2h}^{(1)}(\nu)&s_{2h}^{(2)}(\nu)
\end{array}
\right]^{-1}=\frac{1}{W(\nu)}\left[
\begin{array}{cc}
s_{2h}^{(2)}(\nu)&-s_{2h}^{(2)}(\nu+1)\\
-s_{2h}^{(1)}(\nu)&s_{2h}^{(1)}(\nu+1)
\end{array}
\right]\\
W(\nu)=&s_{2h}^{(2)}(\nu)s_{2h}^{(1)}(\nu+1)-s_{2h}^{(1)}(\nu)s_{2h}^{(2)}(\nu+1)\,.
\end{align}
Let us remind that the Casoratian $W(\nu)$ satisfies the equation
\begin{equation}
p_0(\nu)W(\nu)-p_2(\nu+2)W(\nu+1)=0\,,
\end{equation} 
and, therefore,
\begin{equation}
W(\nu)=\omega(\nu)\frac{4^{-\nu }\Gamma
	\left(\frac{1}{4}-\nu \right) \Gamma
	\left(\frac{3}{4}-\nu \right)}{\sin (\pi  \nu ) \Gamma
	\left(\frac{7}{6}-\nu \right) \Gamma\left(\frac{11}{6}-\nu \right)}\,,
\end{equation}
where $\omega(\nu)$ is some periodic function. Again, using educated guess and \texttt{PSLQ} we find
\begin{equation}
\omega(\nu)\stackrel{1000}{=}\frac{5 \pi\Gamma
	\left(\frac{3}{4}\right)^2 (3-2 \cos (2 \pi 
	\nu ))}{3 \sqrt{3}
	\Gamma \left(\frac{1}{4}\right)^2}\,.
\end{equation}
Note that the only zeros of $W(\nu)$ in the region 
\[\mathcal{D}=\{\nu|\ \Re \nu\leqslant 0\}\] are located in $\pm\frac{\arccos(3/2)}{2\pi}-n=\pm \frac{i}{2\pi}\ln \tfrac12(3+\sqrt5)-n$, where $n=0,1,\ldots$. 

It easy to see then that
\begin{enumerate}
	\item the only singularities of $T(\nu)$ when $\nu\in \mathcal{D}$ are the simple poles at $\nu = \pm \frac{i}{2\pi}\ln \tfrac12(3+\sqrt5)-n$ ($n=0,1,\ldots$),
	\item $s_2(\nu+1)-s_{2ih}(\nu+1)$ is holomorphic when $\nu\in \mathcal D$,
	\item $\left|(e^{-2\pi|\nu|},1)\cdot T(\nu){s_2(\nu+1)-s_{2ih}(\nu+1)\choose s_2(\nu)-s_{2ih}(\nu)}\right|\to 0$ when $\nu\to\pm i\infty$.
	\item $T(\nu){s_2(\nu+1)-s_{2ih}(\nu+1)\choose s_{2}(\nu)-s_{2ih}(\nu)}$ is real-valued when $\nu\in (-\infty,1]$,
	\item the first row of  $T(\nu)$ has zeros at $\nu = -n$ ($n=0,1,\ldots$),
	\item $\left(s_{2h}^{(1)}(\nu),s_{2h}^{(2)}(\nu)\right)\cdot T(\nu)=(0,1)$, in particular, at $\nu=\pm \frac{\arccos\tfrac32}{2\pi}$.
\end{enumerate}
The first three properties secure that
\begin{align*}
\omega_1(\nu)=&a_1+\frac{b_1+c_1 \sin(2\pi\nu)}{3-2\cos(2\pi\nu)}\,,\\ \omega_2(\nu)=&\frac{b_2}{3-2\cos(2\pi\nu)}\,,
\end{align*}
where $a_{1},\,b_{1,2},\,c_{1}$ are some constants. The property \#4 secures that all constants are real. Then the property \#5 leads to the constraint $a_1=-b_1$. Finally, the last property gives the constraints $c_1=0$ and $b_1 s_{2h}^{(1)}\left(\frac{\arccos \frac32}{2\pi}\right)+b_2 s_{2h}^{(2)}\left(\frac{\arccos \frac32}{2\pi}\right)=0$ for $b_1$ and $b_2$. Using \texttt{PSLQ} for the last constraint, we obtain
\begin{align*}
\omega_1(\nu)=&\frac{4a_1\sin^2(\pi\nu)}{3-2\cos(2\pi\nu)}\,,\\ \omega_2(\nu)=&\frac{8\sqrt{\frac23}a_1}{3-2\cos(2\pi\nu)}\,,
\end{align*}
At this point we have only one constant to be fixed. We fix it by explicitly  calculating $S_2$ at $d=1$ to find $S_2(1/2)=\frac{\pi ^{3/2}}{3}$. We obtain
\begin{equation}
	a_1=\frac{3^{3/4} \Gamma \left(\frac{1}{4}\right)}{\sqrt{2} \Gamma \left(-\frac{1}{4}\right)}\,.
\end{equation}
Thus, our result is
\begin{multline}\label{eq:sunrise}
S_2(\nu)= \Gamma \left(\tfrac{4}{3}-\nu \right) \Gamma
\left(\tfrac{3}{2}-\nu \right) \Gamma
\left(\tfrac{5}{3}-\nu \right)\Bigg\{ s_{2ih}(\nu)-\frac{\Gamma \left(\frac{1}{4}\right)^2 3^{3/4}}{2 \pi(3-2\cos(2\pi\nu))}\\
\times\left[s_{2h}^{(1)}(\nu)+\sqrt{\frac83}s_{2h}^{(2)}(\nu)\right]\Bigg\},
\end{multline}
where $s_{2ih}$, $s_{2h}^{(1)}$, and $s_{2h}^{(2)}$ are defined in Eqs.~\eqref{eq:s2ih}, \eqref{eq:s2h1}, and \eqref{eq:s2h2}, respectively.
 
In the next Section we present some numerical results obtained with this representation.

\subsection*{Four-loop watermelon integral}
\begin{figure}[h]
	\unitlength = 0.5mm
	\vspace{5mm} 	    
	\centering
	\begin{fmffile}{mis}
	\parbox{3cm}{
		\centering
		\begin{fmfgraph}(25,40)
			\fmfleft{v1}
			\fmfright{v2}
			\fmftop{v3}
			\fmfbottom{v4}
			\fmf{phantom}{v1,v,v2} 
			\fmf{phantom}{v3,v,v4} 
			\fmf{plain,left=45,tension=0.5}{v,v} 
			\fmf{plain,left=135,tension=0.5}{v,v} 
			\fmf{plain,left=-45,tension=0.5}{v,v} 
			\fmf{plain,left=-135,tension=0.5}{v,v} 
		\end{fmfgraph}\vspace{0mm}\\ 
	\centering $T_1=\Gamma^4(1-\nu)$} \hspace{4mm}
	\parbox{3cm}{
		\centering
		\begin{fmfgraph}(25,40)
			\fmftop{v1}
			\fmfbottom{v2} 
			\fmf{plain}{v1,v2} 
			\fmf{plain,left=1}{v1,v2,v1} 
			\fmf{plain,left=0.5}{v1,v2,v1}
		\end{fmfgraph}\vspace{0mm}\\ 
	\centering $T_4$} \hspace{4mm} 
	\parbox{3cm}{
		\centering
		\begin{fmfgraph}(25,40)
			\fmftop{v1}
			\fmfbottom{v2} 
			\fmf{plain}{v1,va,vb,v2}
			\fmfdot{va,vb} 
			\fmf{plain,left=1}{v1,v2,v1} 
			\fmf{plain,left=0.5}{v1,v2,v1}
		\end{fmfgraph}\vspace{0mm}\\ 
	\centering $T_5$} 
	\end{fmffile}
	\caption{Master integrals. Solid lines correspond to the propagators $1/(k^2-1)$.}\label{fig:mis}	
	\vspace{5mm} 	    
\end{figure}
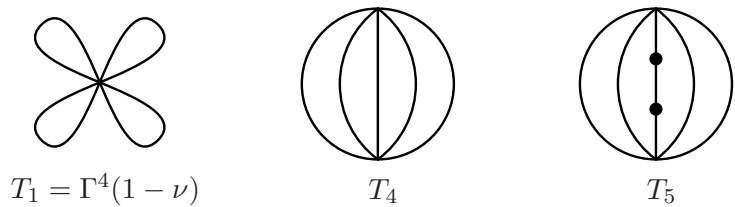

Let us consider the four-loop watermelon tadpole topology. There are three master integrals depicted in Fig.~\ref{fig:mis}.
The first master integral is trivial, $T_1=\Gamma^4(1-\nu)$, and the last two satisfy a coupled system of dimensional recurrence relations:
\begin{align}
T_4(\nu+1)=&\frac{5 \left(435 \nu ^2-674 \nu +263\right)
	T_1(\nu )}{(3 \nu -2)_2 (4 \nu -3)_4}
	+\frac{\left(1629 \nu ^2-4308 \nu
	+2890\right) T_4(\nu )}{(3\nu -2)_2 (4 \nu -3)_4} \nonumber\\
&
	-\frac{1575 T_5(\nu )}{(3 \nu -2)_2(4 \nu -3)_4}\\
T_5(\nu+1)=&\frac{(\nu
	-1) (27 \nu -19) (30 \nu -23) T_1(\nu )}{4(3 \nu -2)_3 (4 \nu
	-3)_2}+\frac{\left(2646 \nu ^3-11403 \nu ^2+13932 \nu
		-5110\right) T_4(\nu )}{20 
		(3 \nu -2)_3 (4 \nu -3)_2} \nonumber\\
	&-\frac{15 (18 \nu -13) T_5(\nu )}{4 \nu 
		(3 \nu -2)_2 (4 \nu -3)_2}\,.
\end{align} 
These equations lead to the second-order recurrence relation for $T_4$:
\begin{multline}\label{eq:T4recurrence}
1125 T_4(\nu )+ (2 \nu -1)_2 \left(549 \nu ^2+16\right)
T_4(\nu +1)
-4 (2 \nu
-1)_4 (3 \nu +1)_2(4 \nu
+1) (4 \nu +3) T_4(\nu +2)\\=5\nu \left(1965 \nu ^3-86 \nu
^2-296 \nu -48\right) \Gamma(-\nu)^4
\end{multline}
We pass to a new function $t_4$ with
\begin{equation}
T_4(\nu)=F(\nu)t_4(\nu)\,,
\end{equation}
where 
\begin{equation}
	F(\nu)=\left(\frac{5}{8}\right)^{\nu } \Gamma (2-2 \nu ) \Gamma \left(\frac{4}{3}-\nu \right) \Gamma \left(\frac{5}{3}-\nu \right)\,.
\end{equation}
Substituting this in Eq.~\eqref{eq:T4recurrence}, we have
\begin{gather}\label{eq:t4recurrence}
\sum_{k=0}^2 p_k(\nu+k)t_4(\nu+k)	= r(\nu)\,,
\end{gather}
where
\begin{align}
	p_0(\nu)=&400 (3 \nu -2) (3 \nu -1)\,,\nonumber\\
	p_1(\nu)=&2 \left(549 \nu ^2-1098 \nu +565\right)\,,\nonumber\\
	p_2(\nu)=&-45 (4 \nu -5) (4 \nu -7)\,,\nonumber\\
	r(\nu) =& 
	-\frac{5 \left(1965 \nu ^3-86 \nu ^2-296 \nu -48\right)\Gamma (-\nu )^4}{(2 \nu -1)F(\nu+1)}
	\,.\label{eq:ps,r}
\end{align}	
The characteristic equation $3600+1098\lambda-720\lambda^2=0$ has the solutions
\begin{equation}
	\lambda_1=\frac{25}8\,,\quad \lambda_2=-\frac85\,.
\end{equation}
The zeros of $p_2(\nu)$ are
\begin{equation}
\nu_1=\frac54\,,\quad \nu_2=\frac74\,.
\end{equation}
So, we search for the solution in the form
\begin{equation}\label{eq:t4sol}
	t_{4h}^{(1)}(\nu)=\lambda_1^{\nu}\sum _{n=0}^{\infty } \left[\frac{\lambda_1^{-\nu_1-n} a_n}{\nu -\nu_1-n}+ \frac{ \lambda_1^{-\nu_2-n} b_n}{\nu -\nu_2 - n}\right]\,,
\end{equation}
where the sequences $\{a_0,a_1,\ldots\}$ and $\{b_0,b_1,\ldots\}$
are determined recursively:
\begin{gather}
	a_{-1}=0,\ a_0=1,\ a_{n\geq 1}=\frac{\left(549 ( 4 n -3)^2 + 256\right) a_{n-1}}{2880 n (2 n-1)}+\frac{5 (12 n-17) (12 n-13) a_{n-2}}{72 n (2 n-1)}\,,\nonumber\\
	\label{eq:seq-ab}
	b_{-1}=0,\ b_{n\geq 1}=\frac{\left(549 (4 n-1)^2 + 256\right) b_{n-1}}{2880 n (2 n+1)}+\frac{5 (12 n-11) (12 n-7) b_{n-2}}{72 n (2 n+1)}\,.
\end{gather}
In order to determine the only non-fixed coefficient $b_0$ we again apply the convergence constraint (similar to the previous example, the polynomial $Q(\nu,\sigma)$ vanishes)
\begin{equation}
\lim_{n\to\infty} n (a_n+b_n) = 0\,.
\end{equation}
from which we have
\begin{equation}
	b_0=-0.7299831875...
\end{equation}
Using some heuristic conjectures and \texttt{PSLQ} algorithm \cite{FergBai1991}, we find
\begin{equation}\label{eq:constant}
b_0\stackrel{1000}{=}-7 \sqrt{\frac{5}{6}} \frac{\Gamma \left(\frac{3}{4}\right)^2}{\Gamma \left(\frac{1}{4}\right)^2}\,.
\end{equation}
Therefore, we have the following homogeneous solution
\begin{equation}
	t_{4h}^{(1)}(\nu)=\sum _{n=0}^{\infty } \left[\left(\frac{25}{8}\right)^{\nu-5/4-n}\frac{a_n}{\nu -5/4-n}-
	c \left(\frac{25}{8}\right)^{\nu-7/4-n}\frac{ b_n}{\nu -7/4 - n}\right]\,,
\end{equation}
where $a_n$ and $b_n$ are defined by Eqs.~\eqref{eq:seq-ab} and \eqref{eq:constant}.

The second homogeneous solution can be found in a similar way by taking $S$ to be the set of zeros of the trailing coefficient $p_0(x)$, i.e., $S=\left\{\frac13,\frac23\right\}$. However, again, it is simpler to use the symmetry of the homogeneous part of Eq.~\eqref{eq:t4recurrence} under the replacement
$t_{4h}(\nu)\to \frac{5^{\nu } \Gamma \left(\frac{7}{4}-\nu \right)t_{4h}(2-\nu)}{\Gamma \left(\frac{5}{3}-\nu \right) \Gamma \left(\frac{4}{3}-\nu \right) \Gamma \left(\nu -\frac{1}{4}\right)}$ followed by $	\nu\to 2-\nu$. Therefore, we may write the second solution as
\begin{equation}
	t_{4h}^{(2)}(\nu)= \frac{5^{\nu } \Gamma \left(\frac{7}{4}-\nu \right)t_{4h}^{(1)}(2-\nu)}{\Gamma \left(\frac{5}{3}-\nu \right) \Gamma \left(\frac{4}{3}-\nu \right) \Gamma \left(\nu -\frac{1}{4}\right)}\,.
\end{equation}
It is easy to check that $t_{4h}^{(2)}(\nu)$ is independent of $t_{4h}^{(1)}(\nu)$. Moreover, using \texttt{PSLQ}, we obtain 
\begin{equation}
	W(\nu)=t_{4h}^{(2)}(\nu)t_{4h}^{(1)}(\nu+1)-t_{4h}^{(1)}(\nu)t_{4h}^{(2)}(\nu+1) = \frac{7\Gamma \left(\frac{3}{4}\right)^2 \left(2-\cos (2 \pi  \nu )\right)5^{\nu}\Gamma \left(\frac{1}{4}-\nu \right) \Gamma \left(\frac{3}{4}-\nu \right)}{4\sqrt{30} \Gamma \left(\frac{1}{4}\right)^2\sin\pi\!  \left(\nu -\frac{3}{4}\right)\Gamma \left(\frac{4}{3}-\nu \right) \Gamma \left(\frac{5}{3}-\nu \right)}
\end{equation}

We write the general solution of \eqref{eq:t4recurrence} in the form 
\begin{equation}\label{eq:general}
t_{4}(\nu) = 
\omega_1(\nu)t_{4h}^{(1)}(\nu)+\omega_2(\nu)t_{4h}^{(2)}(\nu)+ t_{4ih}(\nu)\,,
\end{equation}
where $\omega_{1,2}(\nu)$ are arbitrary periodic functions. The inhomogeneous solution $t_{4ih}(\nu)$ is defined in the same way as in Eqs.~\eqref{eq:s2ih} and \eqref{eq:AR} with $p_k$ and $r$ now defined by Eq.~\eqref{eq:ps,r}. 

The periodic factors can be fixed in a similar way as in the previous example. 
We finally find
\begin{equation}
	\omega_1(\nu)=\frac{16 \left(\frac{2}{15}\right)^{1/4} }{5 \sqrt{\pi }}\Gamma \left(\tfrac{1}{4}\right)^2\,,\quad \omega_2(\nu)=0\,.
\end{equation}
\subsection*{Four-loop cat-eye graph}
\begin{figure}[h]
	\unitlength = 0.5mm
	\vspace{5mm} 	    
	\centering
	\begin{fmffile}{mis1}
		\parbox{3cm}{
			\centering
			\begin{fmfgraph}(15,40)
				\fmftop{v1}
				\fmfbottom{v2}
				\fmfleft{p1} 
				\fmfright{p2}
				\fmf{plain}{p1,p2} 
				\fmf{plain,left=1}{v1,v2,v1} 
				\fmf{plain,left=0.4}{v1,v2,v1}
			\end{fmfgraph}\vspace{1mm}\\ 
			\centering $T_9$} \hspace{4mm} 
		\parbox{3cm}{
			\centering
			\begin{fmfgraph}(40,40)
				\fmftop{v1}
				\fmfbottom{v2}
				\fmfleft{p1} 
				\fmfright{p4}
				\fmf{phantom}{p1,p2}
				\fmf{phantom}{p3,p4}
				\fmf{plain,tension=0.8}{p2,p3}
				\fmfdot{p4} 
				\fmf{plain,left=1}{v1,v2,v1} 
				\fmf{plain,left=0.4}{v1,v2,v1}
			\end{fmfgraph}\vspace{1mm}\\ 
			\centering $T_{10}$}
	\end{fmffile}
	\caption{Master integrals. Solid lines correspond to the propagators $1/(k^2-1)$.}\label{fig:mis1}	
	\vspace{5mm} 	    
\end{figure}
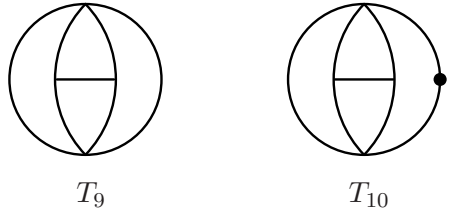

Let us now consider the four-loop cat-eye graph. In the highest sector there are two master integrals depicted in Fig.~\ref{fig:mis1}.
The dimensional recurrence relations for these two master integrals have the form
\begin{equation}\label{eq:T10eq}
\left(\begin{array}{c}
T_9(\nu+1)\\
T_{10}(\nu+1)
\end{array}\right)
-\left(
\begin{array}{cc}
\frac{(2 \nu -3) \left(50 \nu ^2-101 \nu +47\right)}{2(\nu -1) \nu  (2 \nu -1)^2 (3 \nu -4)_3} & -\frac{13 \nu ^2-28 \nu +13}{(\nu -1) \nu  (2 \nu -1)^2 (3 \nu -4)_3} \\
\frac{(2 \nu -3) (13 \nu -17)}{12 (\nu -1)^2 \nu  (2 \nu -1) (3 \nu -4)} & -\frac{5 \nu -7}{6 (\nu -1)^2 \nu  (2 \nu -1) (3 \nu -4)} \\
\end{array}
\right)
\left(\begin{array}{c}
T_9(\nu)\\
T_{10}(\nu)
\end{array}\right) =\ldots
\end{equation}
where the dots in the right-hand side denote the contribution of the simpler masters. The homogeneous part of the second-order recurrence relation for the dotted integral $T_{10}$ has the form
\begin{multline}\label{eq:T10recurrence}
-9 (13 \nu -4) T_{10h}(\nu )-2 (2 \nu -1) \left(455 \nu ^3-1050 \nu ^2+691 \nu -120\right) \nu  T_{10h}(\nu +1)\\
+12 (\nu +1) (2 \nu -1) (2 \nu +1) (3 \nu -2) (3 \nu -1) (13 \nu -17) \nu ^3 T_{10h}(\nu +2)=0\,.
\end{multline}
Passing to a new function $t_{10}$,
\begin{equation}
T_{10}(\nu)=\Gamma (2-2 \nu ) \Gamma (2-\nu ) \Gamma \left(\frac{7}{3}-\nu \right)t_{10}(\nu)\,,
\end{equation}
we have
\begin{gather}\label{eq:t10recurrence}
\sum_{k=0}^2 p_k(\nu+k)t_{10}(\nu+k)= 0\,,
\end{gather}
where
\begin{align}
p_0(\nu)=&-3 (\nu -1) (3 \nu -4) (13 \nu -4)\,,\nonumber\\
p_1(\nu)=&-455 \nu ^3+2415 \nu ^2-4156 \nu +2316\,,\nonumber\\
p_2(\nu)=&9 (\nu -2) (3 \nu -8) (13 \nu -43)\,.\label{eq:ps}
\end{align}	
The characteristic equation $\lambda ^2-\frac{35 \lambda }{27}-\frac{1}{3}=0$ has the solutions
\begin{equation}
\lambda_1=\frac{1}{54} \left(35+13 \sqrt{13}\right)\,,\quad \lambda_2=\frac{1}{54} \left(35-13 \sqrt{13}\right)\,.
\end{equation}
The zeros of $p_2(\nu)$ are
\begin{equation}
\nu_1=2\,,\quad \nu_2=\frac83\,,\quad \nu_3=\frac{43}{13}\,.
\end{equation}
So, according to the prescriptions of the previous section, we search for the solution in the form
\begin{equation}\label{eq:t10sol}
t_{10h}^{(1)}(\nu)=\lambda_1^{\nu}\sum _{n=0}^{\infty } \left[\frac{\lambda_1^{-\nu_1-n} a_n}{\nu -\nu_1-n}+\frac{ \lambda_1^{-\nu_2-n} b_n}{\nu -\nu_2 - n}+\frac{ \lambda_1^{-\nu_3-n} c_n}{\nu -\nu_3 - n}\right]\,,
\end{equation}
where the sequences $\{a_0,a_1,\ldots\}$, $\{b_0,b_1,\ldots\}$, and $\{c_0,c_1,\ldots\}$
are defined recursively:
\begin{gather}
a_{-1}=0,\ \ a_{n\geq 1}=\frac{455 n^3-1050 n^2+691 n-120}{9 n (3 n-2) (13 n-17)}a_{n-1}+\frac{(n-1) (3 n-4) (13 n-4) }{3 n (3 n-2) (13 n-17)}a_{n-2}\,,\nonumber\\
b_{-1}=0,\ \ b_{n\geq 1}=\frac{12285 n^3-3780 n^2-2763 n+238}{81 n (3 n+2) (39 n-25)} b_{n-1}+\frac{(3 n-2) (3 n-1) (39 n+14)}{9 n (3 n+2) (39 n-25)} b_{n-2}\,,\nonumber\\
\label{eq:seq(a,b,c)}
c_{-1}=0,\ \ c_{n\geq 1}=\frac{5915 n^3+9555 n^2+3628 n+72 }{9 n (13 n+17) (39 n+25)}c_{n-1}+\frac{(n+1) (13 n+4) (39 n-1)}{3 n (13 n+17) (39 n+25)} c_{n-2}\,.
\end{gather}
The above recurrence relations determine $a_{n\geqslant0}$, $b_{n\geqslant0}$, and  $c_{n\geqslant0}$ via the starting coefficients $a_0$, $b_0$, and $c_0$, respectively.
In order to determine (up to an overall constant) the starting coefficients, we calculate the polynomial $Q_l(\sigma)$, Eq.~\eqref{eq:Q}. We have
\begin{align}
	Q_0(\sigma)=& -\frac{845}{162} \left(35 \sqrt{13}+169\right) \sigma +\frac{26}{81} \left(1363 \sqrt{13}+6137\right)\,,\\
	Q_1(\sigma)=& -\frac{169}{81}  \left(35 \sqrt{13}+169\right)\,.
\end{align}
Note that $Q_2=0$ as it should be. Thus, we have two conditions,
\begin{align}
	\sum_{n=0}^{\infty} \left\{a_n\lambda_1^{-\nu_1-n}Q_l(\nu_1+n)+b_n\lambda_1^{-\nu_2-n}Q_l(\nu_2+n)+c_n\lambda_1^{-\nu_3-n}Q_l(\nu_3+n)\right\}&=0\quad(l=0,1)\,,
\end{align}
for three coefficients $a_0,\,b_0,\,c_0$. We obtain
\begin{equation}\label{eq:rels}
\frac{b_0}{a_0}=0.38888888\ldots\stackrel{1000}{=}\frac7{18}\,,\quad \frac{c_0}{a_0}=0.000000\ldots\stackrel{1000}{=}0\,.
\end{equation}
Let us remark here that for practical reasons it might be simpler to use another approach to discover numbers in Eq.~\eqref{eq:rels}. Namely, we calculate with high precision the right-hand side of the recurrence relation \eqref{eq:t10recurrence} substituting $t_{10}(\nu)$ with the contribution of each term in square brackets in Eq.~\eqref{eq:t10sol} for two random values of $\nu$. Then we have a linear system of the form
\begin{gather*}
	c_{11}a_0+c_{12}b_0+c_{13}c_0=0\,,\\
	c_{21}a_0+c_{22}b_0+c_{23}c_0=0\,,
\end{gather*}
where $c_{ij}$ are some high-precision numbers. Then the ratios  \eqref{eq:rels} can be obtained from this system.

Therefore, we have the following homogeneous solution
\begin{equation}
t_{10h}^{(1)}(\nu)=\sum _{n=0}^{\infty } \left[\lambda_1^{\nu-2-n}\frac{a_n}{\nu - 2-n}+\lambda_1^{\nu-8/3-n}\frac{ b_n}{\nu -8/3 - n}\right]\,,
\end{equation}
where $\lambda_1=\frac{1}{54} \left(35+13 \sqrt{13}\right)$, $a_0=1$, $b_0=7/18$, and $a_{n>0}$ and $b_{n>0}$ are defined recursively by Eq.~\eqref{eq:seq(a,b,c)}.

The second homogeneous solution can be found in a similar way by taking $S$ to be the set of zeros of the trailing coefficient $p_0(x)$, i.e., $S=\left\{\frac43,1,\frac4{13}\right\}$. We present here only the final result
\begin{equation}
t_{10h}^{(2)}(\nu)=\sin(\pi \nu)\sum _{n=0}^{\infty } \left[(-\lambda_2)^{\nu-4/3+n}\frac{\tilde{a}_n}{\nu-4/3+n}+(-\lambda_2)^{\nu-1+n}\frac{\tilde{b}_n}{\nu-1+n}\right]\,,
\end{equation}
where $\tilde{a}_n$ and  $\tilde{b}_n$ are defined recursively by
\begin{multline}\label{eq:seq(at)}
\tilde{a}_{-1}=0,\ \tilde{a}_{0}=1,\\ \tilde{a}_{n\geq 1}=\frac{12285 n^3-20790 n^2+8577 n-356}{27 n (3 n-1) (39 n-40)}\tilde{a}_{n-1}+\frac{(3 n-4) (3 n-2) (39 n-1)}{n (3 n-1) (39 n-40)}\tilde{a}_{n-2}\,,
\end{multline}
\begin{multline}\label{eq:seq(bt)}
\tilde{b}_{-1}=0,\ \tilde{b}_{0}=4\cdot 3^{1/3},\\ \tilde{b}_{n\geq 1}=\frac{455 n^3-315 n^2-44 n+24}{3 n (3 n+1) (13 n-9)}\tilde{b}_{n-1}+\frac{3 (n-1) (3 n-1) (13 n+4)}{n (3 n+1) (13 n-9)}\tilde{b}_{n-2}\,.
\end{multline}

It is easy to check numerically that $t_{10h}^{(2)}(\nu)$ is independent of $t_{10h}^{(1)}(\nu)$. 
Examining the analytical properties of the integrals $T_{9,10}$ in a similar way as for the previous example, we obtain the final result  
\begin{equation}\label{eq:t10final}
	t_{10}(\nu)=\frac{4 \pi ^2}{\Gamma \left(-\frac{2}{3}\right)}t_{10h}^{(1)}(\nu)+ t_{10ih}(\nu)
\end{equation}
It is important that the inhomogeneous solution  $t_{10ih}(\nu)$ in \eqref{eq:t10final}, which we do not present here, can be constructed automatically using the \texttt{DREAM} package\footnote{Again, we refer the reader to the parallel paper \cite{DREAM} for details.}.

\subsection*{Three-loop box in special kinematics}

\begin{figure}[h]
	\unitlength = 0.5mm
	\vspace{5mm} 	    
	\centering
	\begin{fmffile}{mis2}
		\parbox{3cm}{
			\centering
			\begin{fmfgraph*}(70,30)
				\fmfset{arrow_len}{3mm}
				\fmfset{dot_size}{1mm}
				\fmfleft{l1,l2}
				\fmfright{r1,r2}
				\fmf{fermion,label=$p$}{l1,v1} 
				\fmf{fermion,label=$p$}{l2,v2} 
				\fmf{fermion,label=$p$}{v1a,r1} 
				\fmf{fermion,label=$p$}{v2a,r2}
				\fmf{dashes,tension=0.4}{v1,v1a} 
				\fmf{dashes,tension=0.4}{v2,v2a}
				\fmffreeze
				\fmf{dashes,tension=0.4}{v1,v2a} 
				\fmf{dashes,tension=0.4,rubout}{v2,v1a}
				\fmffreeze
				\fmfdot{v1,v2,v1a,v2a}
				\fmf{plain}{v1,v2}
				\fmf{plain}{v1a,v2a}				 
			\end{fmfgraph*}\vspace{3mm}\\ 
			\centering $\quad J_5$} 
	\end{fmffile}
	\caption{Master integrals. Solid lines correspond to the propagators $1/(k^2-1)$, dashed --- to $1/k^2$. The momentum $p$ is on mass shell, $p^2=1$.}\label{fig:mis2}	
	\vspace{5mm} 	    
\end{figure}
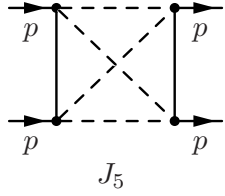

All previous examples related the integrals with massive internal lines. In order to demonstrate the applicability of our method also to the integrals with massless lines, we present here the result for the integral corresponding to the diagram in Fig.~\ref{fig:mis2}. This integral is relevant for the three-loop hard contributions to the energy levels of the weakly bound QED systems, like positronium. There are two master integrals in the highest sector, so the integral without squared propagators satisfies a second-order recurrence relation.
Passing to ${j}_5(\nu)$ connected with the original integral via 
\begin{align}
	{J}_5(\nu)=&\frac{\Gamma (2-\nu )}{\sin (\pi  \nu ) \Gamma \left(\nu -\frac{1}{2}\right) \Gamma \left(\nu -\frac{5}{4}\right)}{j}_5(\nu)\,,
\end{align}
we have the following recurrence relations
\begin{gather}
	3 (\nu -1) (4 \nu -5) {j}_5(\nu )-\left(28 \nu ^2-28 \nu+9\right) {j}_5(\nu +1)+4 \nu  (4 \nu +1) {j}_5(\nu +2) = \ldots\,,
\end{gather}
where dots in the right-hand side denote simpler master integrals.
Using the described approach, we find the following homogeneous solutions for ${j}_5$:
\begin{align}\label{eq:ps5sol}
{j}_{5h}^{(1)}(\nu)=&\sum _{n=0}^{\infty } \left[\frac{a_n}{\nu -7/4-n}
+\frac{b_n}{\nu - 2 - n}\right]\,,\nonumber\\
{j}_{5h}^{(2)}(\nu)=&\left({3}/{4}\right)^{\nu }{j}_{5h}^{(1)}(3-\nu)\,,
\end{align}
where
\begin{gather}
a_{-1}=0,\ a_{0}=1,\ a_{n\geq 1}=\frac{112 n^2-168 n+71}{16 n (4 n-1)}a_{n-1}-\frac{3 (2 n-3) (4 n-5)}{8 n (4 n-1)}a_{n-2}\,,\nonumber\\
b_{-1}=0,\ b_0=-\frac{2 \sqrt{2} \Gamma \left(\frac{3}{4}\right)^4}{\pi ^2},\ b_{n\geq 1}=\frac{28 n^2-28 n+9}{4 n (4 n+1)}b_{n-1}-\frac{3 (n-1) (4 n-5)}{4 n (4 n+1)} b_{n-2}\,.\nonumber\\
\end{gather} 

\section{Computational issues}

First, we would like to note that the representations like \eqref{eq:s2h1} are ideally fitted to obtaining the $\epsilon$ expansion. One should simply expand under the summation sign which amounts to the replacement $(\nu_0-\epsilon-\nu_i-n)^{-1}\to\epsilon^k(\nu_0-\nu_i-n)^{-k-1}$ for $O(\epsilon^k)$ term.

One might question the possibility to obtain high-precision results for the representations like \eqref{eq:s2h1}. Indeed, the sum converges harmonically, as $\sum_n 1/n^2$, and the direct summation of $N$ terms would give only $\log_{10} N$ decimal digits. Fortunately, the convergence acceleration technique described in Refs.~\cite{Broadhurst1996,LeeMingulov:2016:SummerTime} can be successfully applied. In the ancillary file \texttt{Meromorphic.nb}, we present the \emph{Mathematica} procedures for the calculation of all homogeneous solutions obtained in this paper, namely, $s_{2h}^{(1,2)}$, $t_{4h}^{(1,2)}$, $t_{10h}^{(1,2)}$, and $j_{5h}^{(1,2)}$.

\begin{figure}
	\centering
	\includegraphics[]{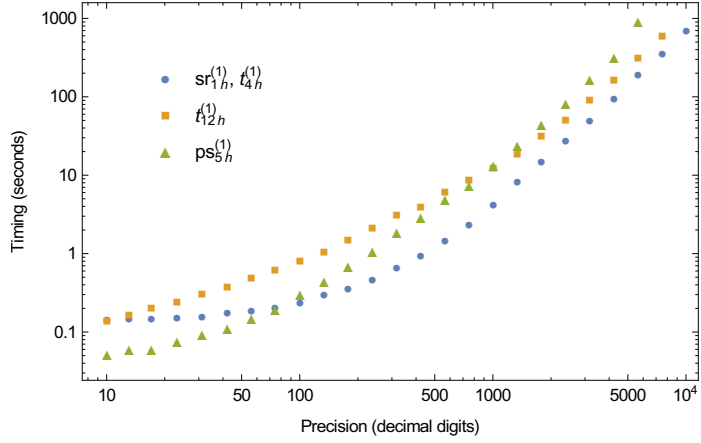}
	\caption{Calculation times of the homogeneous solutions found in the present paper.}\label{fig:efficiency} 
\end{figure}

In Fig.~\ref{fig:efficiency} we present some timing results for the calculation of the homogeneous solutions. As it can be seen, the time of calculation scales as $O(p^\alpha)$, where $p$ is the requested precision and $\alpha\approx 2.3$. The almost quadratic dependence on $p$ could have been anticipated if one takes into account that the computational complexity of the convergence acceleration is quadratic in the number of terms, and the latter is approximately proportional to $p$. 

In a parallel paper \cite{DREAM} we present the \texttt{DREAM} program suitable for the required high-precision numerical computation using the results for the homogeneous solutions obtained in the present paper. Here we will only  demonstrate the effectiveness of our method by presenting the expansions of $S_2$ near $d=3$:
\allowdisplaybreaks
\begin{multline}
\frac{(1-4 \epsilon ) (1-6 \epsilon )S_2(3/2-\epsilon)}{8\Gamma(1/2+\epsilon)^3}=
\frac{1}{\epsilon }-\frac{9 \ln3}{2}-\left(\frac{33
	\text{Li}_2\left(\frac{1}{3}\right)}{4}+\frac{21
	\text{Li}_2\left(\frac{2}{3}\right)}{4}-\frac{11 \pi ^2}{8}-\frac{33}{4} \ln2 \ln
3\right)\epsilon\\
-\left(9
\text{Li}_3\left(\tfrac{1}{3}\right)+18
\text{Li}_3\left(\tfrac{2}{3}\right)-9
\text{Li}_2\left(\tfrac{1}{3}\right) \ln3+26 \zeta (3)-\frac{9 \ln^3 3}{4}+\frac{27}{2} \ln 2 \ln ^23+3
\pi ^2 \ln 3\right)\epsilon ^2\\
-\bigg(-27
\text{Li}_2\left(\tfrac{2}{3}\right){}^2+\frac{93}{8}
\text{Li}_4\left(\tfrac{1}{4}\right)+\frac{189}{2}
\text{Li}_4\left(\tfrac{1}{3}\right)-25
\text{Li}_4\left(\tfrac{1}{2}\right)+189
\text{Li}_4\left(\tfrac{2}{3}\right)+\frac{135}{4}\text{Li}_4\left(\tfrac{3}{4}\right)\\
-\frac{27}{2} \text{Li}_2\left(\tfrac{2}{3}\right)
\ln ^2 3+81
\text{Li}_3\left(\tfrac{1}{3}\right) \ln
3+108 \text{Li}_3\left(\tfrac{2}{3}\right)
\ln 3+\frac{621}{4} \zeta (3) \ln 3-\frac{211 \pi ^4}{360}+\frac{701 \ln^4 2}{24}-\frac{117 \ln ^4 3}{8}\\
-45 \ln^3 2 \ln 3+\frac{27}{2} \ln 2 \ln^3 3-\frac{211}{12} \pi ^2 \ln^2 2+\frac{135}{4} \ln^2 2 \ln^2 3+\frac{45}{2} \pi ^2 \ln 2 \ln 3\bigg)\epsilon^3+O\left(\epsilon
^4\right)
\end{multline}

The high-precision numerical computation of $S_2$ near $d=2$ can be obtained equally easily,
\begin{multline}
\frac{S_2(1-\epsilon)}{\Gamma(1+\epsilon)^3}\stackrel{1000}{=}
-9.10918116586732895891315508535203270601035285811860984353754\ldots\\
+39.6338053538347457129557664099889626125710042446395050863062\ldots \epsilon\\ -148.277921359084179510413643775668953748455629003078326240908\ldots \epsilon^2\\
+503.899909010171775492908647195760723892036963129334838713739\ldots \epsilon^3\\\vdots 
\end{multline}
Unfortunately, the basis of transcendental numbers involved in the above expansion appears to be not quite clear to us\footnote{We thank David Broadhurst for sharing his considerations about the leading term of expansion.} which prohibited the use of \texttt{PSLQ} algorithm.

\section{Conclusion}
In this paper we have formulated an approach to the construction of the  solutions of the higher-order recurrence relations in the form of a sum over poles, with coefficients defined recursively, \eqref{eq:solutionForm}, \eqref{eq:c_recurrence}. 
The main advantage of the described approach is that the analytical properties of the solutions are very clear (the position of poles is explicit, the behavior at infinity can be easily determined). These are exactly the properties that are required for the application of the DRA method.  
Several explicit examples of the application of this technique are given, see Eqs.~\eqref{eq:s2h1}, \eqref{eq:t4sol}, \eqref{eq:t10sol}, and \eqref{eq:ps5sol}.

It is quite remarkable that three out of four examined recurrence relations (and also some more not presented here) exhibited a hidden symmetry $\nu\to a-\nu$ with $a$ equal to $2$ or $3$. The only exclusion is the homogeneous recurrence relation for the cat-eye topology, Eq.~\eqref{eq:t10recurrence}. However for this topology one could speculate that there might exist a linear combination $\tilde{t}_{10}(\nu)=c_0(\nu){t}_{10}(\nu)+c_1(\nu){t}_{10}(\nu+1)$ with coefficients $c_{0,1}(\nu)$ being the rational functions of $\nu$, such that the dimensional recurrence for $\tilde{t}_{10}$ again has the symmetry $\nu\to a-\nu$. In general, a better understanding of the transformations 
\[
f(\nu)\to \tilde{f}(\nu)=\sum_{k=0}^{n-1} c_k(\nu)f(\nu+k) 
\]
is very desirable.
These transformations may lead to essential simplification of the coefficients of the recurrence relations \eqref{eq:recurrence}.

\acknowledgments

R.L. is grateful to Andrei Pomeransky for the interest to the work and useful discussions. This work is supported by the grant of the “Basis” foundation for theoretical physics and by RFBR grant 17-02-00830.

\bibliographystyle{elsarticle-num}

\begin{thebibliography}{10}
	\expandafter\ifx\csname url\endcsname\relax
	\def\url#1{\texttt{#1}}\fi
	\expandafter\ifx\csname urlprefix\endcsname\relax\def\urlprefix{URL }\fi
	\expandafter\ifx\csname href\endcsname\relax
	\def\href#1#2{#2} \def\path#1{#1}\fi
	
	\bibitem{ChetTka1981}
	K.~G. Chetyrkin, F.~V. Tkachov, Integration by parts: {T}he algorithm to
	calculate $\beta$-functions in 4 loops, Nucl.~Phys.~B 192 (1981) 159.
	
	\bibitem{Tkachov1981}
	F.~V. Tkachov, A theorem on analytical calculability of 4-loop renormalization
	group functions, Physics Letters B 100~(1) (1981) 65--68.
	
	\bibitem{Remiddi1997}
	E.~Remiddi, Differential equations for {F}eynman graph amplitudes, Nuovo {C}im.
	A110 (1997) 1435--1452.
	\newblock \href {http://arxiv.org/abs/hep-th/9711188}
	{\path{arXiv:hep-th/9711188}}.
	
	\bibitem{Kotikov1991b}
	A.~V. Kotikov, Differential equation method: {T}he {C}alculation of {N} point
	{F}eynman diagrams, Phys. {L}ett. B267 (1991) 123--127.
	
	\bibitem{Aoyama2015}
	T.~Aoyama, M.~Hayakawa, T.~Kinoshita, M.~Nio, {Tenth-Order Electron Anomalous
		Magnetic Moment --- Contribution of Diagrams without Closed Lepton Loops},
	Phys. Rev. D91~(3) (2015) 033006, [Erratum: Phys. Rev.D96,no.1,019901(2017)].
	\newblock \href {http://arxiv.org/abs/1412.8284} {\path{arXiv:1412.8284}}.
	
	\bibitem{Laporta2017}
	S.~Laporta, {High-precision calculation of the 4-loop contribution to the
		electron g-2 in QED}, Phys. Lett. B772 (2017) 232--238.
	\newblock \href {http://arxiv.org/abs/1704.06996} {\path{arXiv:1704.06996}}.
	
	\bibitem{Laporta2000}
	S.~Laporta, High precision calculation of multiloop {F}eynman integrals by
	difference equations., Int.~J.~Mod.~Phys.~A 15 (2000) 5087.
	
	\bibitem{Tarasov1996}
	O.~V. Tarasov, Connection between {F}eynman integrals having different values
	of the space-time dimension, Phys.~Rev.~D 54 (1996) 6479.
	\newblock \href {http://arxiv.org/abs/hep-th/9606018}
	{\path{arXiv:hep-th/9606018}}.
	
	\bibitem{Lee2010}
	R.~Lee, {Space-time dimensionality D as complex variable: Calculating loop
		integrals using dimensional recurrence relation and analytical properties
		with respect to D}, Nuclear Physics B 830 (2010) 474.
	\newblock \href {http://arxiv.org/abs/0911.0252} {\path{arXiv:0911.0252}}.
	
	\bibitem{LeeMingulov:2016:SummerTime}
	R.~N. Lee, K.~T. Mingulov, {Introducing SummerTime: a package for
		high-precision computation of sums appearing in DRA method}, Computer Physics
	Communications 203 (2016) 255--267.
	
	\bibitem{Lee2010a}
	R.~N. Lee, Calculating multiloop integrals using dimensional recurrence
	relation and {D}-analyticity, Vol. 205-206, 2010, pp. 135--140.
	\newblock \href {http://arxiv.org/abs/1007.2256} {\path{arXiv:1007.2256}}.
	
	\bibitem{LeeSmSm2010a}
	R.~N. Lee, A.~V. Smirnov, V.~A. Smirnov, Dimensional recurrence relations: an
	easy way to evaluate higher orders of expansion in $\epsilon$, Vol. 205-206,
	Elsevier {BV}, 2010, pp. 308--313.
	\newblock \href {http://arxiv.org/abs/1005.0362} {\path{arXiv:1005.0362}}.
	
	\bibitem{LeeSmi2010}
	R.~Lee, V.~Smirnov, {Analytic Epsilon Expansions of Master Integrals
		Corresponding to Massless Three-Loop Form Factors and Three-Loop g-2 up to
		Four-Loop Transcendentality Weight}, J. High Energy Phys. 1102 (2011) 102.
	\newblock \href {http://arxiv.org/abs/1010.1334} {\path{arXiv:1010.1334}}.
	
	\bibitem{LeeSmSm2011}
	R.~N. Lee, A.~V. Smirnov, V.~A. Smirnov, {On Epsilon Expansions of Four-loop
		Non-planar Massless Propagator Diagrams}, Eur. Phys. J. C 71 (2011) 1708.
	\newblock \href {http://arxiv.org/abs/1103.3409} {\path{arXiv:1103.3409}}.
	
	\bibitem{LeeTer2010}
	R.~Lee, I.~Terekhov, Application of the {DRA} method to the calculation of the
	four-loop {QED}-type tadpoles, J. High Energy Phys. 1101 (2011) 068.
	\newblock \href {http://arxiv.org/abs/1010.6117} {\path{arXiv:1010.6117}}.
	
	\bibitem{Lee2011e}
	R.~N. Lee, A.~V. Smirnov, V.~A. Smirnov, {Master Integrals for Four-Loop
		Massless Propagators up to Transcendentality Weight Twelve}, Nucl. Phys. B
	856 (2012) 95--110.
	\newblock \href {http://arxiv.org/abs/1108.0732} {\path{arXiv:1108.0732}}.
	
	\bibitem{LeeSmirnov2012}
	R.~N. Lee, V.~A. Smirnov, {The Dimensional Recurrence and Analyticity Method
		for Multicomponent Master Integrals: Using Unitarity Cuts to Construct
		Homogeneous Solutions}, J. High Energy Phys. 1212 (2012) 104.
	\newblock \href {http://arxiv.org/abs/1209.0339} {\path{arXiv:1209.0339}}.
	
	\bibitem{LeeMarquardSmirnovSmirnovSteinhauser2013}
	R.~N. Lee, P.~Marquard, A.~V. Smirnov, V.~A. Smirnov, M.~Steinhauser,
	{Four-loop corrections with two closed fermion loops to fermion self energies
		and the lepton anomalous magnetic moment}, J. High Energy Phys. 1303 (2013)
	162.
	\newblock \href {http://arxiv.org/abs/1301.6481} {\path{arXiv:1301.6481}}.
	
	\bibitem{LeeSmirnov2016}
	R.~N. Lee, V.~A. Smirnov, {Evaluating the last missing ingredient for the
		three-loop quark static potential by differential equations}, Journal of High
	Energy Physics 1610 (2016) 89.
	\newblock \href {http://arxiv.org/abs/1608.02605} {\path{arXiv:1608.02605}}.
	
	\bibitem{LeeSmirnovSmirnovSteinhauser2016}
	R.~N. Lee, A.~V. Smirnov, V.~A. Smirnov, M.~Steinhauser, {Analytic three-loop
		static potential},\newblock \href {http://arxiv.org/abs/1608.02603}
	{\path{arXiv:1608.02603}}.
	
	\bibitem{DREAM}
	R.~N. Lee, K.~T. Mingulov, {DREAM}, a program for arbitrary-precision
	computation of dimensional recurrence relations solutions, and its
	applications,\newblock \href {http://arxiv.org/abs/1712.05173}
	{\path{arXiv:1712.05173}}.
	
	\bibitem{Tulyakov:2011}
	D.~N. Tulaykov, A procedure for finding asymptotic expansions for solutions of
	difference equations, Proceedings of the Steklov Institute of Mathematics
	272~(2) (2011) 162--167.
	
	\bibitem{FergBai1991}
	H.~R.~P. Ferguson, D.~H. Bailey, {A Polynomial Time, Numerically Stable Integer
		Relation Algorithm}, Tech. rep. (1991).
	
	\bibitem{Broadhurst1996}
	D.~J. Broadhurst, {On the enumeration of irreducible k-fold Euler sums and
		their roles in knot theory and field theory},\newblock \href
	{http://arxiv.org/abs/hep-th/9604128} {\path{arXiv:hep-th/9604128}}.
	
\end{thebibliography}

\end{document}